\documentclass{article}
\usepackage{amssymb}
\usepackage{amsfonts}
\usepackage{amsmath}

\input{tcilatex}

\begin{document}

\title{Ordering ambiguity revisited via position dependent mass
pseudo-momentum operators}
\author{Omar Mustafa$^{1}$ and S.Habib Mazharimousavi$^{2}$ \\
Department of Physics, Eastern Mediterranean University, \\
G Magusa, North Cyprus, Mersin 10,Turkey\\
$^{1}$E-mail: omar.mustafa@emu.edu.tr\\
$^{2}$E-mail: habib.mazhari@emu.edu.tr}
\maketitle

\begin{abstract}
Ordering ambiguity associated with the von Roos position dependent mass
(PDM) Hamiltonian is considered. An affine locally scaled first order
differential introduced, in Eq.(9), as a PDM-pseudo-momentum operator. Upon
intertwining our Hamiltonian, which is the sum of the square of this
operator and the potential function, with the von Roos $d$-dimensional
PDM-Hamiltonian, we observed that the so-called von Roos ambiguity
parameters are strictly determined, but not necessarily unique. Our new
ambiguity parameters' setting is subjected to Dutra's and Almeida's [11]
reliability test and classified as good ordering.

\end{abstract}

\section{Introduction}

Quantum mechanical Hamiltonians with position dependent mass (PDM)
constitute interesting and useful models for the study of many physical
problems [1-31]. They are used in the energy density many-body problem [1],
in the determination of the electronic properties of the semiconductors [2]
and quantum dots [3], in quantum liquids [4], in $^{3}He$ clusters [5] and
metal clusters [6], in the Bohmian approach to quantum theory (cf., e.g.
[7]), in the full and partial wave-packet revivals(cf., e.g., [8]), etc.
Comprehensive reviews on the applicability of such position dependent mass
settings could be found in the sample of references in [1-12]

However, it is concreted that an ordering ambiguity conflict arises in the
process of defining a unique kinetic energy operator, due non-commutativity
between the momentum operator $\hat{p}_{x}=-i\partial _{x}$ and the position
dependent mass $M\left( x\right) =m_{\circ }m\left( x\right) $. A problem
that has shown poor advancement over the last few decades.

In general, working on PDM Hamiltonians is inspired by the von Roos
Hamiltonian [15] proposal (with $\hbar =2m_{\circ }=1$)%
\begin{equation}
H=-\frac{1}{2}\left[ m\left( x\right) ^{\alpha }\partial _{x}m\left(
x\right) ^{\beta }\partial _{x}m\left( x\right) ^{\gamma }+m\left( x\right)
^{\gamma }\partial _{x}m\left( x\right) ^{\beta }\partial _{x}m\left(
x\right) ^{\alpha }\right] +V\left( x\right)
\end{equation}%
subjected to von Roos constraint%
\begin{equation}
\alpha +\beta +\gamma =-1\text{ };\text{ \ }\alpha ,\beta ,\gamma \in 
\mathbb{R}%
\end{equation}%
Hamiltonian (1) may, in a straightforward manner and with the constraint
(2), be very well recast ( cf., e.g., [10]) as%
\begin{equation}
H=-\partial _{x}\left( \frac{1}{m\left( x\right) }\right) \partial _{x}+%
\tilde{V}\left( x\right) ,
\end{equation}%
with%
\begin{equation}
\tilde{V}\left( x\right) =\frac{1}{2}\left( 1+\beta \right) \frac{m^{\prime
\prime }\left( x\right) }{m\left( x\right) ^{2}}-\left[ \alpha \left( \alpha
+\beta +1\right) +\beta +1\right] \frac{m^{\prime }\left( x\right) ^{2}}{%
m\left( x\right) ^{3}}+V\left( x\right) ,
\end{equation}%
where primes denote derivatives. Obviously, nevertheless, the profile of $%
\tilde{V}\left( x\right) $ (namely the first two terms in (4)) changes as
the parameters \ $\alpha ,\beta ,$ and $\gamma $ change, manifesting
therefore the eruption of ordering ambiguity in the process of choosing the
kinetic energy operator $\hat{T}$. Hence, $\alpha ,\beta ,$ and $\gamma $
are usually called the von Roos ambiguity parameters.

Several proposals for the kinetic energy operator are suggested in
literature. Amongst exist; the Gora and Williams ($\beta =\gamma =0,$ $%
\alpha =-1$) [16], Ben Danial and Duke ($\alpha =\gamma =0,$ $\beta =-1$)
[17], Zhu and Kroemer ($\alpha =\gamma =-1/2,$ $\beta =0$) [18], and Li and
Kuhn ($\beta =\gamma =-1/2,$ $\alpha =0$) [19]. However, the Hermiticity of
the kinetic energy operator, the current density conservation, the
experimental results [20-25], and the condensed matter theories [26,27] may
give some ideas on the identity of the von Roos ambiguity parameters.
Applying Hamiltonian (1) to an abrupt heterojunction between two crystals
(cf., e.g., sample of references in [25-28,30]), for example, implied that
for $\alpha \neq \gamma $ the wave function vanishes at the heterojunction
(i.e., the heterojunction plays the role of an impenetrable barrier). Hence,
the only feasible cases are due $\alpha =\gamma $ to ensure the continuity
of $m\left( x\right) ^{\alpha }\psi \left( x\right) $ and $m\left( x\right)
^{\alpha +\beta }\left[ \partial _{x}\psi \left( x\right) \right] $ at the
heterojunction boundary.

Very recently, however, Dutra and Almeida [11] have carried out a
reliability test on the orderings available in literature. They have used an
exactly solvable Morse model and concluded that the orderings of Gora and
Williams ($a=\beta =\gamma =0,$ $\alpha =-1$) [16], and Ben Danial and Duke (%
$a=\alpha =\gamma =0,$ $\beta =-1$) [17] should be discarded for they result
in complex energies. Nevertheless, they have classified the ordering of Zhu
and Kroemer ($a=0,$ $\alpha =\gamma =-1/2,$ $\beta =0$) [18], and that of Li
and Kuhn ($a=\alpha =0,\beta =\gamma =-1/2$) [19] as good orderings. Yet,
they have shown that Weyl (cf., e.g., Borges in [30]) and Li and Kuhn [19]
orderings are equivalent.

Ultimately, therefore, the continuity conditions at the heterojunction
boundaries and Dutra's and Almeida's [11] reliability test single out Zhu
and Kroemer ($a=0,$ $\alpha =\gamma =-1/2,$ $\beta =0$) [18] as good
ordering. This, in effect, inspires our current methodical proposal and
manifests the introduction of a PDM-pseudo-momentum operator which, in turn,
leads to a new good ordering.

On the other hand, within a Liouvillean-type change of variables spiritual
lines, the point canonical transformation (PCT)\ method for Schr\"{o}dinger
equation often mediates a transition between two different effective
potentials. That is, in the PCT settings, one needs the exact solution of a
potential model in a class of shape invariant potentials to form the
so-called \emph{reference/old }potential. The \emph{reference/old }potential
\ along with its exact solution (i.e. eigenvalues and eigenfunctions) is
then mapped into the so-called \emph{target/new }potential\emph{\ ,} hence
exact solution for the \emph{target/new }potential \ is obtained. For more
details on this issue the reader may refer to , e.g., ref. [14].

In this parer, we recollect (in section 2) the $d$-dimensional form of the
von Roos Hamiltonian suggested by Quesne [10] and introduce our
PDM-pseudo-momentum operator. The consequences of such operator's setting on
the von Roos ambiguity parameters are given in the same section. Moreover,
the corresponding $d$-dimensional radial Schr\"{o}dinger Hamiltonian and the
PCT $d$-dimensional mapping are also reported. Our concluding remarks are
given in section 3.

\section{$d$-dimensional von Roos Hamiltonian and PDM-pseudo-momentum
operators}

Quesne in [10] has suggested a general form of von Roos $d$-dimensional PDM
Schr\"{o}dinger equation%
\begin{gather}
\left\{ -\frac{1}{2}\left[ m\left( \mathbf{q}\right) ^{\alpha }\partial
_{j}m\left( \mathbf{q}\right) ^{\beta }\partial _{j}m\left( \mathbf{q}%
\right) ^{\gamma }+m\left( \mathbf{q}\right) ^{\gamma }\partial _{j}m\left( 
\mathbf{q}\right) ^{\beta }\partial _{j}m\left( \mathbf{q}\right) ^{\alpha }%
\right] \right\} \psi \left( \mathbf{q}\right)  \notag \\
+\left\{ V\left( \mathbf{q}\right) -E\right\} \psi \left( \mathbf{q}\right)
=0,
\end{gather}%
\linebreak where $\mathbf{q}=\left( q_{1},q_{2},\cdots ,q_{d}\right) ,$ $%
\partial _{j}=\partial /\partial q_{j},$ $j=1,2,\cdots ,d,$\textbf{\ }$%
m\left( \mathbf{q}\right) $ is the dimensionless form of the mass $M\left( 
\mathbf{q}\right) =m_{\circ }m\left( \mathbf{q}\right) $, $V\left( \mathbf{q}%
\right) $ is the potential function, and summation runs over repeated
indices. In this case, the $d$-dimensional PDM Schr\"{o}dinger Hamiltonian
reads%
\begin{equation}
H=-\partial _{j}\left( \frac{1}{m\left( \mathbf{q}\right) }\right) \partial
_{j}+\tilde{V}\left( \mathbf{q}\right) ,
\end{equation}%
with%
\begin{equation}
\tilde{V}\left( \mathbf{q}\right) =\frac{1}{2}\left( 1+\beta \right) \frac{%
\partial _{j}\partial _{j}m\left( \mathbf{q}\right) }{m\left( \mathbf{q}%
\right) ^{2}}-\left[ \alpha \left( \alpha +\beta +1\right) +\beta +1\right] 
\frac{\left[ \partial _{j}m\left( \mathbf{q}\right) \right] ^{2}}{m\left( 
\mathbf{q}\right) ^{3}}+V\left( \mathbf{q}\right) .
\end{equation}

Let us now consider, for simplicity, quasi-free-particles' setting (i.e., $%
V\left( \mathbf{q}\right) =0$). Then it would be obvious that the
quasi-free-particles' Hamiltonian structure suggests that the kinetic energy
operator 
\begin{equation}
\hat{T}=-\partial _{j}\left( \frac{1}{m\left( \mathbf{q}\right) }\right)
\partial _{j}+\frac{1}{2}\left( 1+\beta \right) \frac{\partial _{j}\partial
_{j}m\left( \mathbf{q}\right) }{m\left( \mathbf{q}\right) ^{2}}-\left[
\alpha \left( \alpha +\beta +1\right) +\beta +1\right] \frac{\left[ \partial
_{j}m\left( \mathbf{q}\right) \right] ^{2}}{m\left( \mathbf{q}\right) ^{3}}
\end{equation}%
may, mathematically speaking, very well be expressed as the square of a
first-order differential vector operator of a general form%
\begin{equation}
\hat{\Pi}_{j}=-i\left\{ F\left( m\left( \mathbf{q}\right) \right) \partial
_{j}+G_{j}\left( m\left( \mathbf{q}\right) \right) \right\} .
\end{equation}%
This would (with $F\left( m\left( \mathbf{q}\right) \right) \equiv F\left( 
\mathbf{q}\right) =F$, $G_{j}\left( m\left( \mathbf{q}\right) \right) \equiv
G_{j}\left( \mathbf{q}\right) =G_{j}$ for simplicity, and $\hat{T}=\hat{\Pi}%
^{2}=\delta _{ij}\hat{\Pi}_{i}\hat{\Pi}_{j}$) imply 
\begin{equation}
\hat{T}=-F^{2}\partial _{j}\partial _{j}\mathbf{-}\left[ F\left( \partial
_{j}F\right) +2FG_{j}\right] \partial _{j}-\left[ F\left( \partial
_{j}G_{j}\right) +G_{j}G_{j}\right] ,
\end{equation}%
If we compare Eq.(10) with (8) we obtain%
\begin{gather}
\text{\ }F\left( \mathbf{q}\right) =\pm \frac{1}{\sqrt{m\left( \mathbf{q}%
\right) }},  \notag \\
\frac{\partial _{j}m\left( \mathbf{q}\right) }{m\left( \mathbf{q}\right) ^{2}%
}=-2F\partial _{j}F=-\left[ F\left( \partial _{j}F\right) +2FG_{j}\right] , 
\notag \\
G_{j}=\frac{\partial _{j}F}{2}\text{ };\text{ }
\end{gather}

The structure of our first-order differentioal operator is therefore clear
and can be cast as%
\begin{equation}
\hat{\Pi}_{j}=-i\left\{ F\left( \mathbf{q}\right) \partial _{j}+\frac{1}{2}%
\left[ \partial _{j}F\left( \mathbf{q}\right) \right] \right\} \text{.}
\end{equation}%
At this point, it should be noted that our operator $\hat{\Pi}$ is Hermitian
and represents the position-dependent-mass generalization of the ordinary
momentum operator $\hat{p}_{j}=-i\partial _{j}$ (i.e., at constant mass
settings $M\left( x\right) =m_{\circ }$). Hence, $\hat{\Pi}$ could be
labeled, hereinafter, as a \emph{PDM-pseudo-momentum operator}.

\subsection{Consequences of our \emph{PDM-pseudo-momentum operator} $\hat{\Pi%
}$ on the von Roos ambiguity parameters}

In a straightforward manner it is easy to show that%
\begin{equation}
\hat{\Pi}^{2}=-\partial _{j}\left( \frac{1}{m\left( \mathbf{q}\right) }%
\right) \partial _{j}+\frac{1}{4}\frac{\partial _{j}\partial _{j}m(\mathbf{q}%
)}{m(\mathbf{q})^{2}}-\frac{7}{16}\frac{\left[ \partial _{j}m\left( \mathbf{q%
}\right) \right] ^{2}}{m(\mathbf{q})^{3}}.
\end{equation}%
Comparing this result with the kinetic energy operator $T$ in (8) we obtain%
\begin{equation}
\left( 1+\beta \right) =\frac{1}{2}\text{, \ }\left[ \alpha \left( \alpha
+\beta +1\right) +\beta +1\right] =\frac{7}{16},
\end{equation}%
which in turn suggests that the von Roos ambiguity parameters are strictly
determined (but not necessarily unique) as 
\begin{equation}
\beta =-\frac{1}{2}\text{ , and }\alpha =\gamma =-\frac{1}{4}.
\end{equation}%
Hence, the $d$-dimensional von Roos Hamiltonian reads%
\begin{equation}
H=-\partial _{j}\left( \frac{1}{m\left( \mathbf{q}\right) }\right) \partial
_{j}+\tilde{V}(\mathbf{q})=-\vec{\nabla}_{d}\left( \frac{1}{m\left( \mathbf{q%
}\right) }\right) \cdot \vec{\nabla}_{d}+\tilde{V}(\mathbf{q})
\end{equation}%
with%
\begin{equation}
\tilde{V}(\mathbf{q})=\frac{1}{4}\frac{\partial _{j}\partial _{j}m(\mathbf{q}%
)}{m(\mathbf{q})^{2}}-\frac{7}{16}\frac{\left[ \partial _{j}m\left( \mathbf{q%
}\right) \right] ^{2}}{m(\mathbf{q})^{3}}+V(\mathbf{q}).
\end{equation}

At this point, one may wish to subject such ambiguity parameters' settings
(15) to Dutra's and Almeida's [11] reliability test on the exactly solvable\
one-dimensional Morse model (see equations (10)-(16) in [11]) . Such test
shows that the ambiguous term $\nu \left( \alpha ,\beta ,\gamma ,a\right) =%
\sqrt{1/4-2q/c^{2}}=1/4$ (i.e., equation(16) in [11] for $\alpha =\gamma
=-1/4,\beta =-1/2$) and classifies our ordering as a \emph{good-ordering}
(along with \ that of Zhu's and Kroemer's [18].

\subsection{Corresponding $d$-dimensional radial Schr\"{o}dinger Hamiltonian}

We, in the forthcoming developments, shall assume the radial symmetrization
of $m\left( \mathbf{q}\right) $ and $V\left( \mathbf{q}\right) $ in the $d$%
-dimensional radially symmetric Schr\"{o}dinger Hamiltonian for $d\geq 2$.
Under these settings, (17) and (18) imply%
\begin{equation}
H_{r,d}=-\partial _{j}\left( \frac{1}{m\left( r\right) }\right) \partial
_{j}+\tilde{V}(r)=-\vec{\nabla}_{d}\left( \frac{1}{m\left( r\right) }\right)
\cdot \vec{\nabla}_{d}+\tilde{V}(r)
\end{equation}%
where%
\begin{equation}
\tilde{V}(r)=\frac{1}{4}\frac{\partial _{r}^{2}m(r)}{m(r)^{2}}-\frac{7}{16}%
\frac{\left[ \partial _{r}m(r)\right] ^{2}}{m(r)^{3}}+V(r)\,;\text{ \ \ }%
\mathbb{R}
\ni r\in \left( 0,\infty \right)
\end{equation}%
Recollecting that the $d$-dimensional wave function for radially symmetric
Schr\"{o}dinger equation is given by 
\begin{equation}
\Psi \left( \vec{r}\right) =r^{-\left( d-1\right) /2}R_{n_{r},\ell
_{d}}\left( r\right) Y_{\ell _{d},m_{d}}\left( \theta _{1},\theta
_{2},\cdots ,\theta _{d-2},\varphi \right) .
\end{equation}%
would, in turn, when substituted in 
\begin{equation}
\left\{ -\vec{\nabla}_{d}\left( \frac{1}{m\left( r\right) }\right) \cdot 
\vec{\nabla}_{d}+\tilde{V}(r)\right\} \Psi \left( \vec{r}\right) =E_{d}\Psi
\left( \vec{r}\right) ,
\end{equation}%
results in the following $d$-dimensional radial Schr\"{o}dinger equation%
\begin{equation}
\left\{ \frac{d^{2}}{dr^{2}}-\frac{\ell _{d}\left( \ell _{d}+1\right) }{r^{2}%
}+\frac{m^{\prime }\left( r\right) }{m\left( r\right) }\left( \frac{d-1}{2r}-%
\frac{d}{dr}\right) -m\left( r\right) \left[ \tilde{V}\left( r\right) -E_{d}%
\right] \right\} R_{n_{r},\ell _{d}}\left( r\right) =0.
\end{equation}%
Where $\ell _{d}=\ell +\left( d-3\right) /2$ for $d\geq 2,$ $\ell $ is the
regular angular momentum quantum number, and $n_{r}=0,1,2,\cdots $ is the
radial quantum number. Of course, equation (23) is privileged with the
inter-dimensional degeneracies associated with the isomorphism between
angular momentum $\ell $ and dimensionality $d$. On the other hand,
moreover, the $d=1$ (with $%
\mathbb{R}
\ni r\in \left( 0,\infty \right) \longrightarrow 
\mathbb{R}
\ni x\in \left( -\infty ,\infty \right) $) case can be obtained through the
trivial substitutions $\ell _{d}=-1$ and $\ell _{d}=0$ \thinspace for even
and odd parity, $\mathcal{P=}\left( -1\right) ^{\ell _{d}+1}$, respectively.
Yet, a unique isomorphism exists between the $S$-wave ($\ell =0$) energy
spectrum in 3D and in 1D. On this issue, the reader may wish to refer to,
e.g., Mustafa and Znojil [32],\ and Mustafa and Mazharimousavi [13,14,31]
and references cited therein.

\subsection{Corresponding PCT $d$-dimensional mapping}

In this section, we closely follow Mustafa's and Mazharimousavi's recipe
discussed in [14]. Where, a substitution of the form $R\left( r\right)
=m\left( r\right) ^{1/4}\,\phi \left( Z\left( r\right) \right) $ in (22)
would result in $Z^{\prime }\left( r\right) =\sqrt{m\left( r\right) }$,
manifested by the requirement of a vanishing coefficient of the first-order
derivative of $\phi \left( Z\left( r\right) \right) $ ( hence a
one-dimensional form of Schr\"{o}dinger equation is achieved), and suggests
the following point canonical transformation%
\begin{equation}
Z\left( r\right) =\int^{r}\sqrt{m\left( y\right) }dy\text{ }\implies \phi
_{n_{r},\ell _{d}}\left( Z\left( r\right) \right) =m\left( r\right)
^{-1/4}R_{n_{r},\ell _{d}}\left( r\right) .
\end{equation}%
Which in effect implies%
\begin{equation}
\left\{ -\frac{d^{2}}{dZ^{2}}+\frac{\ell _{d}\left( \ell _{d}+1\right) }{%
r^{2}m\left( r\right) }+V_{eff}\left( r\right) -E_{d}\right\} \phi
_{n_{r},\ell _{d}}\left( Z\right) =0,
\end{equation}%
where%
\begin{equation}
V_{eff}\left( r\right) =V\left( r\right) -U_{d}\left( r\right) ;\text{ \ \ }%
U_{d}\left( r\right) =\frac{m^{\prime }\left( r\right) \left( d-1\right) }{%
2r\,m\left( r\right) ^{2}}.
\end{equation}%
It should be noted, however, that the definition of $U_{d}\left( r\right) $
in (25) is now more simplified than that in Eq. (8) of Mustafa and
Mazharimousavi in [14].

On the other hand, an exactly solvable (including conditionally-exactly or
quasi-exactly solvable) $d$-dimensional time-independent \thinspace radial
Schr\"{o}dinger \thinspace wave equation (with a constant mass $M\left(
x\right) =m_{\circ }$ and $\hbar =2m_{\circ }=1$ units)%
\begin{equation}
\left\{ -\frac{d^{2}}{dZ^{2}}+\frac{\mathcal{L}_{d}\left( \mathcal{L}%
_{d}+1\right) }{Z^{2}}+V\left( Z\right) -\varepsilon \right\} \psi
_{n_{r},\ell _{d}}\left( Z\right) =0
\end{equation}%
would form a \emph{reference} for the exact solvability of the \emph{target}
equation (24). That is, if the exact/conditionally-exact/quasi-exact
solution (analytical/numerical) of (26) is known one can construct the
exact/conditionally-exact/quasi-exact solution of (24) through the relation 
\begin{equation}
\frac{\ell _{d}\left( \ell _{d}+1\right) }{r^{2}m\left( r\right) }+V\left(
r\right) -U_{d}\left( r\right) -E_{d}\iff \frac{\mathcal{L}_{d}\left( 
\mathcal{L}_{d}+1\right) }{Z^{2}}+V\left( Z\right) -\varepsilon ,
\end{equation}%
Where $\mathcal{L}_{d}$ is the $d$-dimensional angular momentum quantum
number of the \emph{reference} Schr\"{o}dinger equation. The \emph{reference}
-\emph{target} map is therefore complete and an explicit correspondence (cf.
e.g., Znojil and L\'{e}vai [33] and Mustafa and Mazharimousavi [13,14])
between two bound state problems is obtained.

A power-law position dependent mass of the form $m\left( r\right) =\varsigma
r^{\upsilon }$ , for example, would imply a PCT function 
\begin{equation}
Z\left( r\right) =\sqrt{\varsigma }\int^{r}y^{\upsilon /2}dy=\frac{2\sqrt{%
\varsigma }}{\left( \upsilon +2\right) }\,r^{\left( \upsilon +2\right)
/2}\implies \frac{\left( \upsilon +2\right) }{2}Z\left( r\right) =r\,\sqrt{%
m\left( r\right) }\text{\thinspace }
\end{equation}%
and hence equation (26) gives%
\begin{equation}
U_{d}\left( r\right) =\frac{\upsilon \left( d-1\right) }{2r^{2}m\left(
r\right) }\equiv \frac{2\upsilon \left( d-1\right) }{\left( \upsilon
+2\right) ^{2}Z\left( r\right) ^{2}};\text{ \ }\upsilon \neq -2
\end{equation}%
Relation (28) in effect reads, 
\begin{equation}
\frac{\lambda \left( \lambda +1\right) }{r^{2}m\left( r\right) }\left( \frac{%
\upsilon }{2}+1\right) ^{2}+V\left( r\right) -E_{d}\iff \frac{\mathcal{L}%
_{d}\left( \mathcal{L}_{d}+1\right) }{Z^{2}}+V\left( Z\right) -\varepsilon ,
\end{equation}%
with 
\begin{equation}
\lambda =-\frac{1}{2}+\left\vert \upsilon +2\right\vert ^{-1}\sqrt{\left( 
\frac{\upsilon }{2}+1\right) ^{2}+4\ell _{d}\left( \ell _{d}+1\right)
+2\upsilon \left( 1-d\right) };\text{ \ }\upsilon \neq -2.
\end{equation}%
However, for $\upsilon =-2$ $\Longrightarrow m\left( r\right) =\alpha r^{-2}$
\ we obtain%
\begin{equation}
Z\left( r\right) =\sqrt{\varsigma }\int^{r}t^{-1}dt=\sqrt{\varsigma }\ln r,
\end{equation}%
and hence%
\begin{eqnarray}
\tilde{U}_{d}\left( \upsilon =-2\right) &=&\frac{\ell _{d}\left( \ell
_{d}+1\right) }{\varsigma }-U_{d}\left( r,\upsilon =-2\right)  \notag \\
&=&\frac{\ell _{d}\left( \ell _{d}+1\right) +d-1}{\varsigma }.
\end{eqnarray}%
Which would only add a constant to the left-hand-side of (28) to yield, with 
$\mathcal{L}_{d}=0$ and/or $\mathcal{L}_{d}=-1$ (i.e., only s-states and/or $%
d=1$ states are available from the right-hand-side of (28) ),%
\begin{equation}
V\left( r\right) +\tilde{U}_{d}\left( \upsilon =-2\right) -E_{d}\iff V\left(
q\right) -\varepsilon .
\end{equation}

\section{Concluding Remarks}

In this paper we have developed a Hermitian PDM-pseudo-momentum operator $%
\hat{\Pi}_{j}=-i\left\{ F\left( \mathbf{q}\right) \partial _{j}+\left[
\partial _{j}F\left( \mathbf{q}\right) \right] /2\right\} ,$where $F\left( 
\mathbf{q}\right) =\pm 1/\sqrt{m\left( \mathbf{q}\right) }$. Hereby, the
notion of \emph{PDM-pseudo-momentum operator} is inspired by the fact that $%
\hat{\Pi}$ has an in-built regular momentum operator $\hat{p}_{j}=-i\partial
_{j}$, which is recoverable at constant mass settings (i.e., $M\left( 
\mathbf{q}\right) =m_{\circ }\Longrightarrow F\left( \mathbf{q}\right) =\pm
1 $). Moreover, we have constructed our $d$-dimensional PDM-Hamiltonian, $%
H_{MM}=\hat{\Pi}^{2}+V\left( \mathbf{q}\right) .$

On the other hand, upon intertwining our Hamiltonian, $H_{MM}$, with the von
Roos $d$-dimensional PDM-Hamiltonian, $H_{vR}=\hat{T}_{d}\left( \alpha
,\beta ,\gamma \right) +V\left( \mathbf{q}\right) $, (cf.,e.g., Quesne
[10]), we have observed that the so-called von Roos ambiguity parameters
(i.e., $\alpha ,\beta $ and $\gamma $) are strictly determined (i.e., $\beta
=-1/2$ and $\alpha =\gamma =-1/4$), but not necessarily unique of course.
Therefore, the von Roos $d$-dimensional PDM-Hamiltonian collapses into%
\begin{equation}
H_{vR}\Longrightarrow H_{MM}=-m\left( \mathbf{q}\right) ^{-1/4}\partial
_{j}m\left( \mathbf{q}\right) ^{-1/2}\partial _{j}m\left( \mathbf{q}\right)
^{-1/4}+V\left( \mathbf{q}\right) .
\end{equation}

On the logistical supportive sides of our strict determination of the von
Roos ambiguity parameters $\beta =-1/2$ and $\alpha =\gamma =-1/4$, we
recollect that Bagchi et al [29], while analyzing the so-called
quasi-free-particle problem, have used an intertwining relationship $\eta
H=H_{1}\eta $ (where $\eta $ is a Darbouxal first-order intertwining
operator) and reported that such choices of the ambiguity parameters
correspond to smooth mass functions $m\left( x\right) $ that signalled the
formation of bound states. Yet, Ko\c{c} et al. [30] have started with $%
\alpha =\gamma =0$ and $\beta =-1$ with constant potential $V\left( z\right)
=V_{\circ }$ (equation (3) of Koc et al in [30]) in their study of
transmission probabilities of the scattering problem through a square well
potential with PDM barrier. However,they were forced to change the potential
form (equation (4) of Koc et al in [30]) into%
\begin{equation}
V\left( z\right) =V_{\circ }+\frac{\hbar ^{2}}{8m\left( z\right) ^{2}}\left(
m^{\prime \prime }\left( z\right) -\frac{7m^{\prime }\left( z\right) ^{2}}{%
4m\left( z\right) }\right)
\end{equation}%
which is exactly the same form of the effective potential that comes out
from our eq.(17) with the new $\beta =-1/2$ and $\alpha =\gamma =-1/4$
parameters setting (of course one should mind the units used in this paper, $%
\hbar =2m_{\circ }=1$). Moreover, Dutra's and Almeida's [11] reliability
test resulted in classifying our ordering as a \emph{good-ordering} (along
with \ that of Zhu's and Kroemer's [18], and Li's and Kuhn's [19]).

Therefore, the continuity conditions at the heterojunction boundaries and
Dutra's and Almeida's [11] reliability test would ultimately single out Zhu
and Kroemer ($a=0,$ $\alpha =\gamma =-1/2,$ $\beta =0$) [18] and our new
ordering ($\beta =-1/2,$ $\alpha =\gamma =-1/4$) as \emph{good orderings.}

On the least consequential research stimulant side, such ambiguity
parameters' setting would, in effect, flourish a production-line for new
sets of exactly-solvable, quasi-exactly solvable, and conditionally-exactly
solvable \emph{target/new} Hamiltonian models. The point canonical
transformation (PCT)\ method used in this work exemplifies one of the
methods that generate such spectrum of exact-solvability. For example, for a 
\emph{reference/old} exactly-solvable 
\begin{equation*}
\tilde{V}_{eff}\left( Z\right) =\frac{\mathcal{L}_{d}\left( \mathcal{L}%
_{d}+1\right) }{Z^{2}}+V\left( Z\right)
\end{equation*}%
in (30) there is a corresponding \emph{target/new} exactly-solvable%
\begin{equation*}
\tilde{V}_{eff}\left( r\right) =\frac{\lambda \left( \lambda +1\right) }{%
r^{2}m\left( r\right) }\left( \frac{\upsilon }{2}+1\right) ^{2}+V\left(
r\right)
\end{equation*}%
where $\lambda $ is given by (31) and $\upsilon \neq -2$. Yet, a
comprehensive number of illustrative examples on the generalized
d-dimensional PCT is given by Mustafa and Mazharimousavi in [14]. Of course
other methods designed to obtain exact-solvability do exist. Amongst, we may
name the Lie algebraic method (cf., e.g., L\'{e}vai in [29], intertwining
operators related to supersymmetric quantum mechanics (SUSYQM) method (cf.,
e.g., Quesne in [10]), and the shape-invariance technique (cf., e.g., Quesne
in [10] and Cooper et al. in [29]).

On the feasible applicability side of our strictly determined von Roos
ambiguity parameters, the applicability of such ambiguity parameters' recipe
should not only be restricted to Hermitian PDM Hamiltonians but also to a
broader class of non-Hermitian PDM  $\eta $\emph{-weak-pseudo-Hermitian}
Hamiltonians (cf., e.g., Mustafa and Mazharimousavi [31] and related
references cited therein).\bigskip 

\bigskip \textbf{Acknowledgement:} We would like to thank the referee for
the valuable comments and suggestions.

\end{document}